\documentclass[showpacs,showkeys,amsmath,amssymb,amsfonts]{revtex4}%
\usepackage{graphicx}
\usepackage[usenames]{color}
\usepackage{ulem} %\sout{strike out}
\usepackage{cancel}%\cancelto{0}{x}
\usepackage{epstopdf} %needed to use .eps %might need option --shell-escape (oder -shell-escape?) in old versions
\usepackage{extarrows}

\providecommand{\rbr}[1]{\left( #1 \right)}%
\providecommand{\sqbr}[1]{\left[ #1 \right]} %

\begin{document}

\title{Misusing the entropy maximization in the jungle of generalized entropies}

\author{Thomas Oikonomou}
\email{thomas.oikonomou@nu.edu.kz}
\affiliation{Department of Physics, Nazarbayev University, 53 Kabanbay Batyr Avenue, Astana 010000, Kazakhstan}

\author{G. Baris Bagci}
\email{gbb0002@hotmail.com} \affiliation{Department of Materials Science and Nanotechnology Engineering, TOBB University of Economics and Technology, 06560 Ankara, Turkey}


\begin{abstract}
It is well-known that the partition function can consistently be factorized from the canonical equilibrium distribution obtained through the maximization of the Shannon entropy. We show that such a normalized and factorized equilibrium distribution is warranted if and only if the entropy measure $I \{(p)\}$ has an additive slope i.e. $\partial I \{(p)\} / \partial p_i$ when the ordinary linear averaging scheme is used. Therefore, we conclude that the maximum entropy principle of Jaynes should not be used for the justification of the partition functions and the concomitant thermodynamic observables for generalized entropies with non-additive slope subject to linear constraints. Finally, Tsallis and R\'enyi entropies are shown not to yield such factorized canonical-like distributions.
\end{abstract}

\eid{ }
\date{\today }
\startpage{1}
\endpage{1}
\maketitle

%\newpage
%==================================================
\section{Introduction}
%==================================================
%

The Maximum Entropy (MaxEnt) principle introduced by Jaynes is an information-theoretical approach \cite{Jaynes1}, based on the maximization of the entropy measure $I (\{ p \})$ subject to some relevant constraints. MaxEnt picks the probability distribution with the highest entropy $I (\{ p \})$ from among all the distributions compatible with the given constraints. Then, the obtained distribution is regarded as the best one to describe the available information of the system. To this aim, Jaynes, using the information-theoretic Shannon measure, showed how the equilibrium distributions of the statistical mechanics can be obtained through MaxEnt approach \cite{Jaynes1}, establishing a connection between information theory and statistical mechanics.

Later on, the MaxEnt principle has also been the usual tool for the generalized entropies such as the Tsallis \cite{Tsallis1988} and R\'enyi \cite{Renyi} entropies. These generalized entropies yielded the inverse-power law distributions through the maximization procedure \cite{Bashkirov,TsallisMP,Bagci,Oik2007}. Hence, the MaxEnt approach has mainly been considered as the justification of the equilibrium distributions of these generalized entropies. This led to the numerous applications of these generalized entropies with the aim of a novel non-equilibrium statistical mechanics \cite{Ponmurugan,saberian,rate,Chang,reis,mendes,rotator,Van1,Van2}. Moreover, different approaches equivalent to MaxEnt have been developed as well, to derive the equilibrium distributions \cite{PlastPlastCurado,PlastinoCudaro,
OikBagTir,miller}. However, the applications and a great deal of progress withstanding, we note that there are still some important issues concerning generalized entropies such as the validity of the third law \cite{third}, discrete-continuum transition \cite{Abe2010}, Lesche stability \cite{Lesche,Abe,Lutsko} and incompatibility with the Bayesian updating procedure \cite{Presse}.

Note however that the equilibrium distributions obtained from the MaxEnt principle through the Shannon measure can always be written as a normalized and factorized expression mimicking the ordinary canonical equilibrium distribution $\exp(- \beta \varepsilon _i)/\sum_k \exp(- \beta \varepsilon _k)$. If this would not be the case, one cannot attribute the normalization factor as the partition function, and therefore cannot secure the foundation of the thermodynamic observables, since the partition function is in general essential to the calculation of the thermodynamic observables.

In this work, we investigate whether this essential feature of factorization holds as it does in the case of Shannon entropy. In Section \ref{MaxEntF}, we show that the necessary and sufficient condition for a structurally normalized and factorized equilibrium distribution is the entropy measure $I (\{p\})$ used in the MaxEnt approach to have an additive slope $\partial I (\{ p\}) / \partial p_i$. The Shannon, Tsallis and R\'enyi entropies are then investigated both theoretically and numerically in Section \ref{MaxEntF2}. The concluding remarks are finally presented.

%==================================================
\section{MaxEnt and the Additive Slope: General Formalism}\label{MaxEntF}
%==================================================
%
%
The MaxEnt approach of Jaynes is based on the Lagrange multipliers method, which yields $n$ mathematically independent variables $\{p_1,\ldots,p_n\}=:\{p\}$ by subjecting the entropy measure $I(\{p\})$ to some set of suitable constraints. Then, accordingly, one determines the possible maxima of the aforementioned measure. The notation $:=$ is used for the mathematical definitions. 

In order to obtain the canonical distribution, the usual constraints are the normalization and the mean value on a set $\{\varepsilon_i\}$, $\sum_{i=1}^{n} p_i=1$ and  $U=\sum_{i=1}^{n} p_i \varepsilon_i$, respectively, where we also assume the micro-energies are ordered i.e. $\varepsilon_1<\varepsilon_2<\cdots<\varepsilon_n$.
%
%Moreover, the measure $I(\{p_i\})$ is supposed to be concave.
%
Then, the functional $\Lambda=\Lambda(\{p\},\gamma,\beta)$ to be maximized is given as
\begin{eqnarray}\label{ExtFun}
\Lambda=I(\{p\})
-\gamma \sqbr{\sum_{i=1}^{n} p_i-1}
-\beta \sqbr{\sum_{i=1}^{n} p_i\varepsilon_i-U}\,.
\end{eqnarray}
The factors $\gamma$ and $\beta$ are the so-called Lagrange multipliers associated to the normalization  and the mean value constraints, respectively.
Through the condition $\delta\Lambda=0$, we obtain the following $n+2$-equations
%
%\begin{subequations}\label{eq:03}
\begin{eqnarray}\label{eq:03a}
f(p_i)\big|_{i=1,\ldots,n}&=&\alpha+\beta \varepsilon_i\,,\\
\label{eq:03b}
1&=&\sum_{i=1}^{n} p_i\,,\\
\label{eq:03c}
U&=&\sum_{i=1}^{n} p_i \varepsilon_i\,,
\end{eqnarray}
%\end{subequations}
%
where $\partial I(\{p\})/\partial p_i=f(p_i)+c$ ($c$ is a constant) and $\alpha :=\gamma-c$.
From here on, the function $f (x)$ is called the slope of the entropy measure $I(\{p\})$. 
From Eq. (\ref{eq:03a}), the bijectivity of the function $f$ yields
\begin{equation}\label{MaxEntDistr}
p_i =f^{-1}\rbr{\alpha + \beta \varepsilon_i}\,,
\end{equation}
where $f^{-1} (x)$ is the inverse of the function $f (x)$ such that $f^{-1}(f(x))=x$. Substituting the expression above into Eqs. (\ref{eq:03b}) and (\ref{eq:03c}), we now have
\begin{equation}\label{eq:04}
\sum_{i=1}^{n}f^{-1}(\alpha +\beta \varepsilon_i)=1\qquad\mathrm{and}\qquad
\sum_{i=1}^{n}f^{-1}(\alpha +\beta \varepsilon_i)\varepsilon_i=U\,.
\end{equation}
If the system is solvable, then there exist  Lagrange multipliers $\alpha^\star$ and $\beta^\star$ satisfying Eq. (\ref{eq:04}). 
The starred expressions denote the quantities maximizing the functional $\Lambda(\{p\},\gamma,\beta)$ in Eq. (\ref{ExtFun}).

Assuming the fulfilment of Eq. (\ref{eq:04}), implying that $p_i^\star=f^{-1}(\alpha^\star + \beta^\star \varepsilon_i)$ is normalized,  the question remains as to whether $p_i^\star$ can be written in the following form
\begin{equation}\label{ProbStr}
p^\star_i=f^{-1}(\alpha^\star+\beta^\star \varepsilon_i) = \frac{f^{-1}(\beta^\star\varepsilon_i)}{\sum_{k=1}^{n} f^{-1}(\beta^\star\varepsilon_k)} =: P_i^\star\,,
\end{equation}
for $\alpha^\star\neq0$ and $\beta^\star\neq0$. The above transition from $p_i^\star$ to $P_i^\star$ has always been assumed in the literature without exception to the best of our knowledge. In other words, the distribution $p_i^\star$, once obtained, has always been assumed to be cast into the form $P_i^\star$. 
Note that the structure of $P_i^\star$ is essential for theoretical calculations, since the normalization factor $Z:=\sum_k f^{-1}(\beta^\star\varepsilon_k)$ is identified as the partition function from which the thermodynamic observables are determined. Moreover, $P_i^\star$ warrants the normalization of $p_i^\star$ in large systems \cite{note}.
We now proceed to prove that the relation represented by Eq. (\ref{ProbStr}) is valid if and only if the inverse function $f^{-1} \left( x \right) $ is multiplicative
\begin{equation}\label{multiplicative}
f^{-1}\rbr{\sum_{r=1}^{m} x_r} = \prod_{r=1}^{m}f^{-1}(x_r)\,.
\end{equation}
In order to do this, we first assume that the distribution maximizing the functional in Eq. (\ref{ExtFun}) is given as $p_i^\star=f^{-1}(\alpha^\star+\beta^\star\varepsilon_i)$ in Eq. (\ref{ProbStr}) and the multiplicativity in Eq. (\ref{multiplicative}) hold. Then, $p_i^\star$ can be written as
\begin{equation}\label{multiplicative1}
p_i^\star=f^{-1}(\alpha^\star+\beta^\star\varepsilon_i)
=f^{-1}(\alpha^\star)f^{-1}(\beta^\star\varepsilon_i)\,.
\end{equation}
Hence, the normalization constraint $\sum_{i=1}^{n}p_i^\star=1$, already assumed as a result of Eq. (\ref{ExtFun}), yields
\begin{equation}\label{multiplicative2}
\sum_{i=1}^{n}f^{-1}(\alpha^\star)f^{-1}(\beta^\star\varepsilon_i)=1\,.
\end{equation}
Since $\alpha^\star$ is just the Lagrange multiplier apart from some constants (see below Eq. (\ref{eq:03c})), it is independent of the summation index so that one has
\begin{equation}\label{multiplicative3}
f^{-1}(\alpha^\star)\sum_{i=1}^{n} f^{-1}(\beta^\star\varepsilon_i)=f^{-1}(\alpha^\star)\,Z=1\,,
\end{equation}
which immediately yields
\begin{equation}\label{multiplicative5}
f^{-1}(\alpha^\star)=\frac{1}{Z}
\quad\Rightarrow\quad
\alpha^\star=f\big(Z^{-1}\big)\,.
\end{equation}
%
%By applying the function on both sides of the equality above, we obtain
%
%\begin{equation}\label{multiplicative5}
%\alpha^\star=f\big(Z^{-1}\big)\,.
%\end{equation}
%
Substituting then $\alpha^\star$ above into Eq. (\ref{multiplicative3}), we finally see that $p^\star_i$ can be written as $P_i^\star= f^{-1}(\beta^\star\varepsilon_i)/Z$ thereby confirming Eq. (\ref{ProbStr}).

In order to proceed with the proof in the converse direction, we first show that the multiplicativity of the inverse function $f^{-1} \left( x \right)$ is the same as the additivity of the function $f \left( x \right)$. To do so, we assume that the multiplicativity of the inverse function holds in accordance with Eq. (\ref{multiplicative}). Then, we apply the function $f$ on both sides of Eq. (\ref{multiplicative}) by also  substituting $x_i\rightarrow f(x_i)$ so that one has
\begin{equation}\label{additive}
f\rbr{\prod_{r=1}^{m}x_r}=\sum_{r=1}^{m}f(x_r)\,,
\end{equation} 
which represents the additivity of the function $f$. 
This point can be further clarified by considering the ordinary (natural) logarithm as a function i.e., $f (x) = \ln (x)$ for example. The logarithmic function is additive, since $\ln (x_1 x_2 x_3) = \ln (x_1) + \ln (x_2) + \ln (x_3)$ as a particular case of Eq. (\ref{additive}) for $m = 3$. Its inverse, the exponential function $f^{-1} = \exp (x)$, is then multiplicative i.e. $\exp (x_1 + x_2 + x_3) = \exp (x_1)  \exp (x_2) \exp (x_3)$ conforming to Eq. (\ref{multiplicative}).

To prove the converse direction of our statement, namely if $p_i^\star=P_i^\star$ then $f^{-1}$ is multiplicative,  we apply the function $f$ on both sides of Eq. (\ref{ProbStr}) obtaining
\begin{equation}\label{eq:07b}
\alpha^\star=f\rbr{P_i^\star}-\beta^\star \varepsilon_i\,.
\end{equation}
Since $\alpha$ does not depend on the index $i$, the only option for Eq. (\ref{eq:07b}) to hold is $f$ to be additive. Indeed, in this case, the $i$ dependence vanishes and we again obtain Eq. (\ref{multiplicative5}). To sum up, the passage from the equilibrium distribution $p^\star_i = f^{-1}(\alpha^\star+\beta^\star \varepsilon_i)$ to $P^\star_i= f^{-1}(\beta^\star\varepsilon_i)/Z$ and vice versa is possible if and  only if the function $f(x)$ is additive (see Eq. (\ref{additive})) or equivalently if its inverse $f^{-1} (x)$ is multiplicative (see Eq. (\ref{multiplicative})). Therefore, we conclude that the transition $p_i^\star \longleftrightarrow P_i^\star$ in Eq. (\ref{ProbStr}) with $\alpha^\star\neq0$ and $\beta^\star\neq0$ holds if and only if the entropic measure $I(\{p\})$ has an additive slope $f(p_i)$, yielding the Lagrange multiplier  in Eq. (\ref{multiplicative5}) within the MaxEnt approach.

One may claim, that $\alpha^\star$ can be still extracted from Eq. (\ref{eq:07b}),  even for a nonadditive $f(x)$, by multiplying the former equation with the term $P^\star_i$ and then summing over the index $i$, obtaining
\begin{equation}\label{eq:08}
\widetilde{\alpha} :=\sum_{i=1}^{n} P^\star_i
\big[f(P^\star_i)-\beta^\star\varepsilon_i\big]
=\sum_{i=1}^{n} P^\star_i
f(P^\star_i)-\beta^\star\widetilde{U}\,,
\end{equation}
with $\widetilde{\alpha}=\alpha^\star$ and $\widetilde{U}=U$, where $\widetilde{U}:=\sum_{i=1}^{n}P_i^\star\varepsilon_i$.
However, this move is again dependent on the explicit use of Eq. (\ref{eq:07b}) which is itself well-grounded only when the function $f(x)$ is additive as proven above, implying $\widetilde{\alpha}\neq \alpha^\star$ and $\widetilde{U}\neq U$. In fact, it can be shown that Eq. (\ref{eq:08}) reduces to Eq. (\ref{multiplicative5}) whenever the function $f(x)$ is additive, or equivalently, its inverse $f^{-1} (x)$ is multiplicative. To see that this is indeed so, consider Eq. (\ref{eq:08}) and substitute $P_i^\star = f^{-1}(\beta^\star\varepsilon_i)/Z$ so that one has  
\begin{equation}\label{misc1}
\widetilde{\alpha}= \sum_{i=1}^{n} \frac{f^{-1}(\beta^\star\varepsilon_i)}{Z} \sqbr{f\rbr{\frac{f^{-1}(\beta^\star\varepsilon_i)}{ Z}} - \beta^\star\varepsilon_i}\,.
\end{equation}
Now, assuming that $f(x)$ is additive, we can write 
\begin{eqnarray}\label{misc2}
\nonumber
f\rbr{\frac{f^{-1}(\beta^\star\varepsilon_i)}{Z}}
&=&
f\rbr{f^{-1}(\beta^\star\varepsilon_i)} + 
f(Z^{-1})\\
&=&
\beta^\star\varepsilon_i + 
f(Z^{-1})\,.
\end{eqnarray}
The substitution of the above equation into Eq. (\ref{misc1}) yields 
\begin{equation}\label{misc3}
\widetilde{\alpha}= 
\sum_{i=1}^{n} \frac{f^{-1}(\beta^\star\varepsilon_i)}{Z} f(Z^{-1})
= f(Z^{-1})=\alpha^\star\,.
\end{equation}
Therefore, Eq. (\ref{eq:08}) reduces to Eq. (\ref{multiplicative5}) only if the function $f(x)$ is additive.

Finally, one might consider overcoming the aforementioned obstacle of factorization by modifying the Lagrange multiplier related to the mean energy constraint as $\beta\rightarrow\alpha\zeta\beta$, but by keeping the linear averaging procedure intact. 
Note that the parameter $\zeta$ can have explicit dependence on the particular deformation parameters of the generalized entropy under scrutiny.
Through this modification, one then obtains $p_i^\star=f^{-1}[\alpha^\star(1 + \zeta \beta^\star \varepsilon_i)]$. For the distribution $p_i^\star$ to warrant the factorization, it should satisfy $f^{-1}\Big(\prod_{r=1}^{m}x_r\Big)=\prod_{r=1}^{m}f^{-1}(x_r)$ so that it can yield
\begin{equation}
p_i^\star=\frac{f^{-1}(1 + \zeta \beta^\star \varepsilon_i)}{\sum_{k=1}^{n} f^{-1}(1 + \zeta  \beta^\star \varepsilon_k)}\,.
\end{equation}
The requirement $f^{-1}\Big(\prod_{r=1}^{m}x_r\Big) = \prod_{r=1}^{m}f^{-1}(x_r)$ together with the continuity of $f^{-1} (x)$ uniquely determine the inverse function to be of the form $f^{-1}(x)=x^a$ (see the proof of this particular issue in the Appendix in Ref. \cite{hanel}). Taking also into account that $p_i^\star$ should recover the exponential decay in the ordinary Boltzmann-Gibbs-Shannon limit, there is solely one option left for $p_i^\star$ 
\begin{eqnarray}\label{coupledCon2}
p_i^\star=\frac{(1 + \zeta \beta^\star \varepsilon_i)^{-1/\zeta}}{\sum_{k=1}^{n} (1 + \zeta \beta^\star \varepsilon_k)^{-1/\zeta} }\,,
\end{eqnarray}
so that $\lim_{\zeta\rightarrow0}p_i^\star\sim e^{-\beta^\star\varepsilon_i}$. In other words, the limit $\zeta\rightarrow0$ is required for the distribution to yield ordinary exponential as a limiting case. On the other hand, to recover the ordinary Shannon functional in the maximization procedure, one should have $\alpha\zeta \rightarrow 1$, since only then one has the ordinary internal energy Lagrange multiplier $\beta$. The two limits $\zeta\rightarrow0$ and $\alpha\zeta \rightarrow 1$ necessary to recover the ordinary Boltzmann-Gibbs-Shannon case contradict one another unless the limit $\alpha \rightarrow \infty$ is realized. However, the limit $\alpha \rightarrow \infty$ is untenable, since it implies that the ordinary canonical partition function is zero independent of the micro-states. This point can be explicitly illustrated by inspecting the maximization functional $\Phi$ in the seminal paper of Tsallis \cite{Tsallis1988} which reads
\begin{equation}\label{sample1}
\Phi = I_\mathrm{T}(\{p_i\}) +\alpha \sqbr{\sum_{i=1}^{n} p_i-1} +\alpha \beta (1-q) \sqbr{\sum_{i=1}^{n} p_i\varepsilon_i-U}\,,
\end{equation}
where $I_\mathrm{T}(\{p_i\})$ is the Tsallis entropy and $\zeta= 1-q$. To obtain the ordinary canonical distribution one uses the limit $\zeta\rightarrow0$ (i.e., $q\rightarrow 1$) so that one simultaneously has to have $\alpha\zeta \rightarrow 1$ (i.e., $\alpha\ (1-q) \rightarrow 1$). However, this limit instead becomes $\alpha\zeta \rightarrow 0$ (i.e. $\alpha\ (1-q) \rightarrow 0$) as $q \rightarrow 1$, implying that the ordinary canonical distribution in the limit $q \rightarrow 1$ is obtained at the expense of a micro-canonical maximization (note that the third term on the right hand side of Eq. (\ref{sample1}) becomes zero, leaving only the normalization constraint when $q \rightarrow 1$). The only way out is that one can have $\alpha \rightarrow \infty$ as $q \rightarrow 1$, which renders the partition function of the ordinary canonical distribution zero in general which is nonsensical.

%==================================================
\section{Shannon, Tsallis and R\'enyi Entropies}\label{MaxEntF2}
%==================================================
%
We now consider Shannon, Tsallis and R\'enyi entropy measures below 
%
%\begin{subequations}\label{entropies}
\begin{eqnarray}
\label{entropies_S}
I_\mathrm{S}(\{p\})&=&\sum_{i=1}^{n} p_i \ln(1/p_i)\,,\\
\label{entropies_T}
I_\mathrm{T}(\{p\})&=&\sum_{i=1}^{n} p_i \ln_q(1/p_i)\,,\\
\label{entropies_R}
I_\mathrm{R}(\{p\})&=& \frac{1}{1-q} \ln \left( \sum_{i=1}^{n} p_i^q \right)  \,,
\end{eqnarray}
%\end{subequations}
%
\noindent respectively. The deformed $q$-logarithm is defined as $\ln_q (x) := (x^{1-q}-1)/(1-q)$. It is well-known that Shannon and R\'enyi entropies are additive whereas the Tsallis entropy is non-additive regarding the probabilistic independence. Having maximized them in accordance with Eq. (\ref{ExtFun}), we see that the associated functions $f (x_i)$ and their inverse functions $f^{-1} (x_i)$ can be identified as 
%
%\begin{subequations}\label{Slopes}
\begin{eqnarray}
\label{Slope_a}
f_\mathrm{S}(x_i)=\ln(1/x_i)\,,\hskip1.0cm 
f^{-1}_\mathrm{S}(x_i)=\frac{1}{\exp(x_i)}\,,\hskip1.02cm\\
\label{Slope_b}
f_\mathrm{T}(x_i)= q\ln_q(1/x_i)\,,\hskip0.6cm f^{-1}_\mathrm{T}(x_i)
=\frac{1}{\exp_q(x_i/q)}\,,\hskip0.55cm\\
\label{Slope_c}
f_\mathrm{R}(x_i)= Q\ln_q(1/x_i)\,,\hskip0.5cm
f^{-1}_\mathrm{R}(x_i) =\frac{1}{\exp_q(x_i/Q)}\,,\hskip0.5cm
\end{eqnarray}
%\end{subequations}
%
where the $q$-exponential is defined as $\exp_q(x) := \left[ 1 + \left( 1-q \right)x  \right]^{\frac{1}{1-q}}$ and $Q:= q/\sum_k p_k^q$. A quick inspection of the above functions and their inverses indicates that the only entropy with an additive slope is the Shannon measure due to the natural logarithm. The respective maximized probability  $p_i^\star$ and the normalized measure $P_i^\star$ are denoted as
\begin{eqnarray}\label{MaxProb}
p^\star_{\mathrm{\lambda},i}=f_\mathrm{\lambda}^{-1}( \alpha^\star_\mathrm{\lambda} +\beta^\star_\mathrm{\lambda} \varepsilon_i)\,,\qquad 
P^\star_{\mathrm{\lambda},i}= 
\frac{\displaystyle f_\mathrm{\lambda}^{-1}( \beta^\star_\mathrm{\lambda} \varepsilon_i)}{\displaystyle \sum_k f_\mathrm{\lambda}^{-1}(\beta^\star_\mathrm{\lambda} \varepsilon_k)}\,,
\end{eqnarray}
where  $\lambda=\mathrm{S,T,R}$, $Q^\star=q/\sum_k (p_k^\star)^q$, $\alpha_\lambda^\star=\gamma_\lambda^\star-c^\star_\lambda$ with  $c^\star_\mathrm{S,T}=-1$ and $c^\star_\mathrm{R}=Q^\star/(1-q)$.

We further consider a three-level system $\{\varepsilon_1,\varepsilon_2,\varepsilon_3\}$ with $\varepsilon_1<U<\varepsilon_3$.
The MaxEnt Lagrange multipliers $\alpha^\star_\mathrm{\lambda}$ and $\beta^\star_\mathrm{\lambda}$ are calculated numerically from the Eqs. (\ref{eq:03a})-(\ref{eq:03c}) by virtue of Eqs. (\ref{Slope_a})-(\ref{Slope_c}) for the randomly chosen set $\{\varepsilon_1, \varepsilon_2, \varepsilon_3,U\} = \{1,3,5,2.35\}$. 
The respective values for $\widetilde{\alpha}_{\lambda}$ are calculated from from Eqs. (\ref{eq:08}) and (\ref{MaxProb}).

\begin{figure}[h]
\centering
\includegraphics[height=7cm,width=10cm]{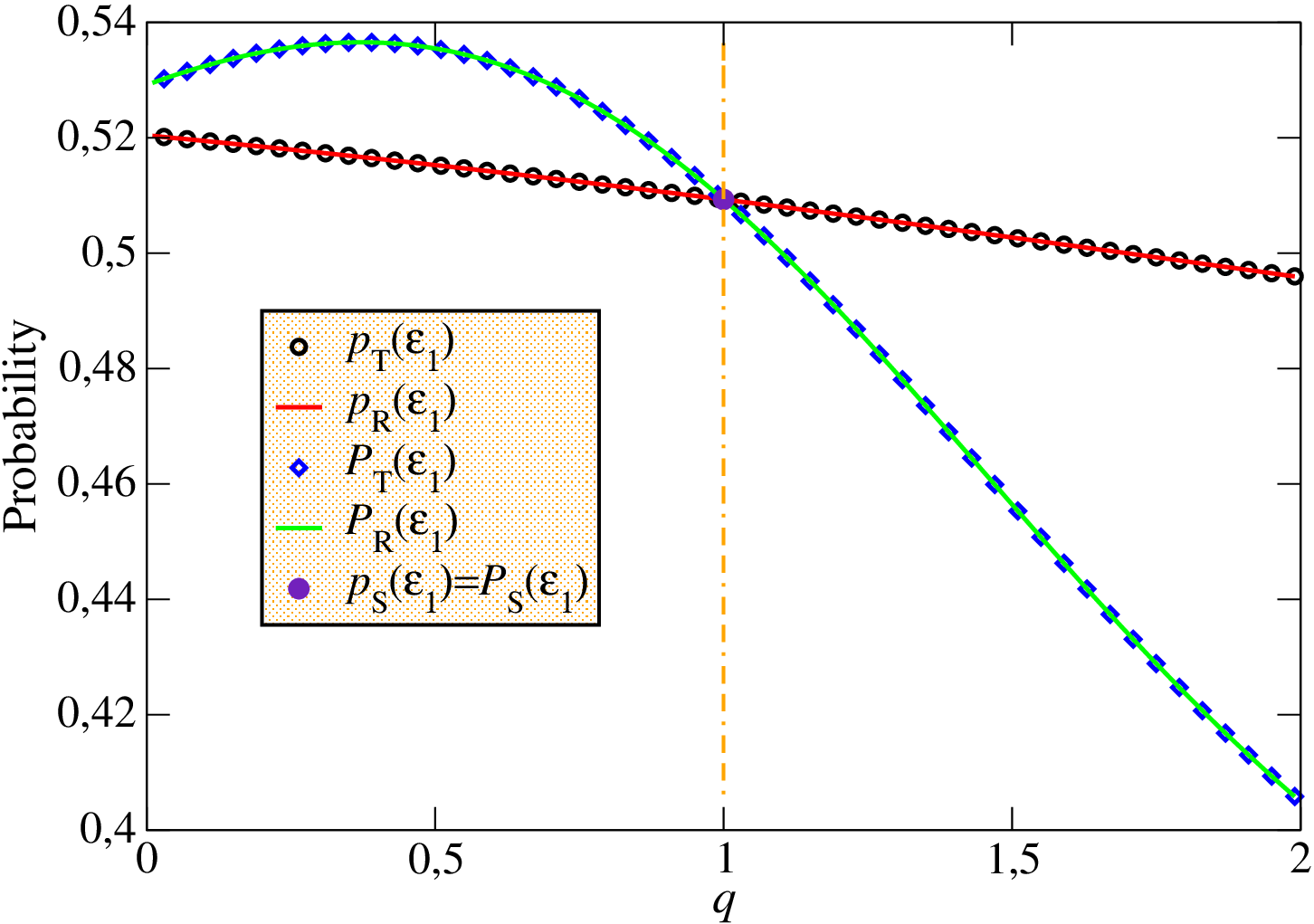}
\caption{The MaxEnt probability  $p^\star_{\mathrm{\lambda},1}$ and the normalized measure $P^\star_{\mathrm{\lambda},1}$ for the state $\varepsilon_1$ with respect to the index $q$ for the Shannon (purple filled circle), Tsallis (black circles and blue diamonds, respectively) and R\'enyi (red and green solid lines, respectively) measures ($\lambda=$S,T,R). The plot demonstrates $p^\star_{\mathrm{\lambda},1}\neq P^\star_{\mathrm{\lambda},1}$ for $q\neq 1$ in accordance with our theoretical results.}\label{Fig1}
\end{figure}

In Fig. \ref{Fig1} we present the plot of  $p^\star_{\lambda,1}$ and $P^\star_{\lambda,1}$ for the state $\varepsilon_1$ with respect to the  index $q\in(0,2)$. The values of $q$ are calculated in steps of $\Delta q=0.01$.
Regarding the Shannon measure, which does not depend on $q$, we have recorded one singe value corresponding to $q=1$. As can be seen, in this case $p^\star_{\mathrm{S},1}$  is equal to $P^\star_{\mathrm{S},1}$ (purple filled circle), which is expected, since both Shannon measure and its slope are additive as a result of (natural) logarithmic dependence.

Regarding the Tsallis measure in the relevant $q$-interval $q\in(0,2)$  \cite{third}, one observes that $p^\star_{\mathrm{T},1}$ (black circles) and $P^\star_{\mathrm{T},1}$ (blue diamonds) coincide only for $q=1$, which is the limiting value for the Tsallis entropy to converge to the Shannon measure. That $p^\star_{\mathrm{T},1}\neq P^\star_{\mathrm{T},1}$ for $q\neq1$ is in accordance with our theoretical prediction, since the slope of the Tsallis measure is non-additive as well as the measure itself.

The results for the R\'enyi measure are exactly the same with the ones of the Tsallis measure for both  $p^\star_{\mathrm{R},1}$ (red solid  line) and $P^\star_{\mathrm{R},1}$ (green solid line). This is to be expected, since both measures are related monotonically to each others yielding the same maxima with $\alpha^\star_\mathrm{T} = \sum_k(p_k^\star)^q\alpha^\star_\mathrm{R}$ and $\beta^\star_\mathrm{T}=\sum_k(p_k^\star)^q \beta^\star_\mathrm{R}$. However, this case is particularly important for our analysis, since the R\'enyi entropy itself is additive while its slope is non-additive. This result shows that it is not the entropy measure itself that counts, but its slope in accordance with our theoretical prediction. In this sense, the Shannon entropy is unique in that both itself and its slope is additive thereby being the perfect point of departure for the canonical distribution in the MaxEnt approach.  

\begin{figure}[h]
\centering
\includegraphics[height=7cm,width=10cm]{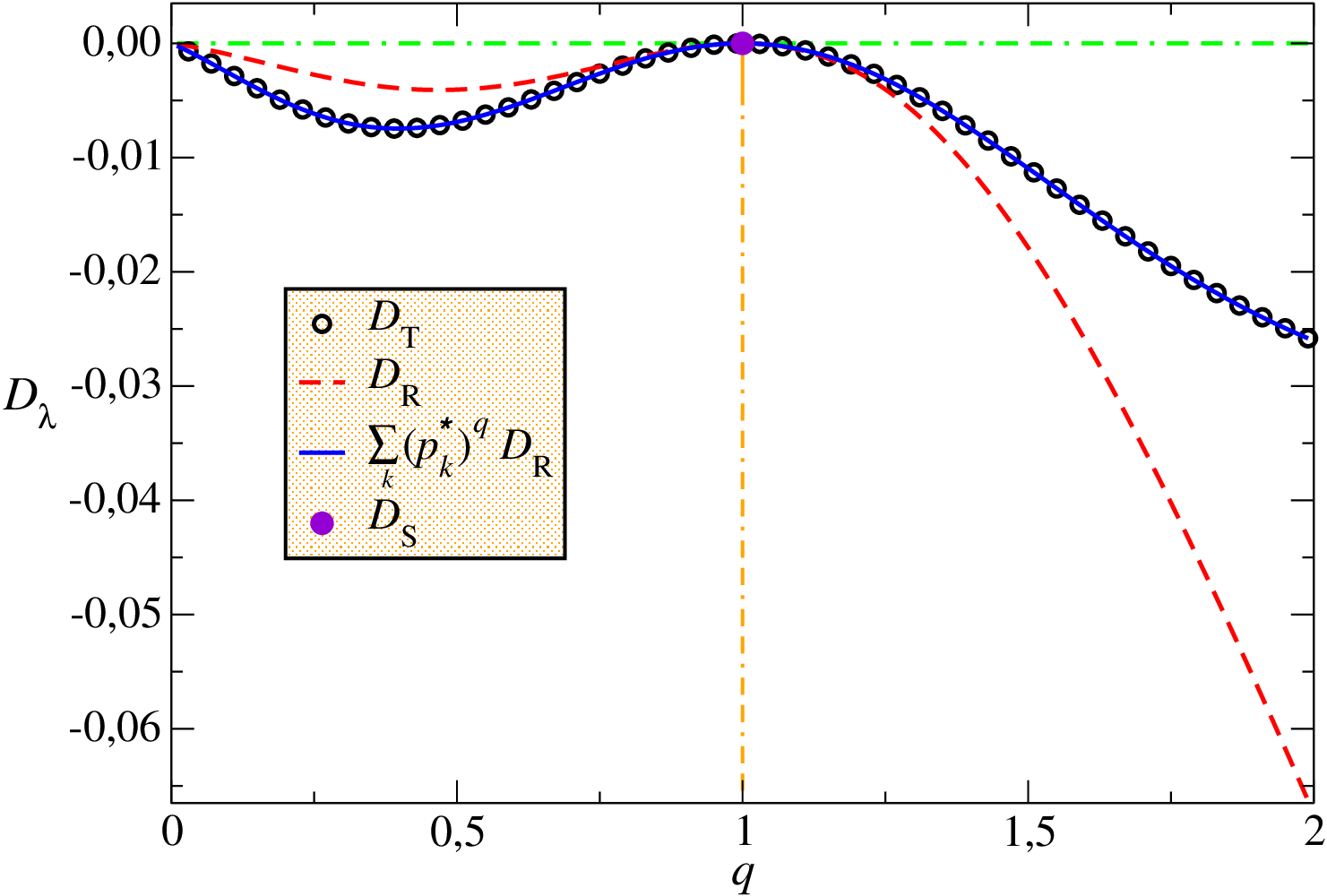}
\caption{The difference $D_\lambda:=\alpha_\lambda^\star - \widetilde{\alpha}_\lambda$ for the Shannon (blue filled circle), Tsallis (black circles) and R\'enyi (red solid line) measures ($\lambda=\mathrm{S,T,R}$) as a function of the index $q$. Note that $\widetilde{\alpha}_\lambda$ reproduces the correct values of the MaxEnt Lagrange multiplier $\alpha^\star_\lambda$ only for $q=1$, which corresponds to the Shannon measure. The blue solid line represents the expression $\sum_{k}(p_k^\star)^q D_\mathrm{R}$ and verifies our analytically obtained relation $D_\mathrm{T}=\sum_{k}(p_k^\star)^q D_\mathrm{R}$.}\label{Fig2}
\end{figure}

We also present the difference $D_\lambda = \alpha_{\lambda}^\star - \widetilde{\alpha}_{\lambda}$ with respect to the index $q\in(0,2)$ in Fig. \ref{Fig2}.
It can be observed that the value of $\widetilde{\alpha}_\mathrm{T,R}$ is equal to the actual MaxEnt Lagrange multiplier $\alpha^\star_\mathrm{T,R}$ (numerically determined as previously) only for $q=1$. This is exactly the point where the slope of both measures becomes additive recovering the Shannon expression with $D_\mathrm{S}=0$ (filled purple circle). 
Note that the numerical difference between $D_\mathrm{T}$ and $D_\mathrm{R}$ (black circles and red dashed line, respectively) is in agreement with our  theoretical prediction $D_\mathrm{T}=\sum_k (p_k^\star)^q D_\mathrm{R}$ (blue solid line)\cite{footnote}.
%

%\newpage
%==================================================
\section{Conclusions}\label{Conclusion}
%==================================================
%
We have shown that the MaxEnt approach applied on an information-theoretic measure $I( \{p \}) $, subject to the normalization and the linear mean value constraints, can yield optimal factorized probability distributions written as $p^\star_i= f^{-1}(x_i)/\sum_k f^{-1}(x_k)$, if and only if the function $f \left( x \right) $ is additive (or equivalently its inverse $f^{-1} \left( x \right)$ being multiplicative). As a well-known example, the Shannon entropy measure in Eq. (\ref{entropies_S}) can be given, since one then has $f(x)=-\ln(x)$ and $f^{-1}(x)=e^{-x}$, yielding the well-known canonical form $p_i^\star= e^{-x_i}/\sum_k e^{-x_k}$.

This result should not be confused with the additivity of the entropy measure itself. For example, R\'enyi entropy is additive as an entropy measure just like the Shannon entropy, but its slope is not additive unlike the Shannon entropy. Therefore, one cannot use the R\'enyi measure and the entropy maximization together in order to obtain a factorized normalized distribution of the form $p^\star_i= f^{-1}(x_i)/\sum_k f^{-1}(x_k)$. As a matter of fact, the MaxEnt distribution may still be obtained satisfying the constraints under consideration even for the entropy measures with the non-additive slope, yet it cannot be expressed in terms of the above normalized and factorized structure $p_i^\star$.
In other words, in this case  one cannot warrant structurally the normalization of $p_i^\star$, since the Lagrange multiplier related to the normalization constraint cannot be separated due to the non-multiplicativity of  $f^{-1} \left( x \right)$.

An immediate consequence of this observation is related to the partition functions. When one has the factorized normalized structure $p^\star_i= f^{-1}(x_i)/\sum_k f^{-1}(x_k)$ as a result of the MaxEnt principle, the denominator $\sum_k f^{-1}(x_k)$ plays the role of the partition function from which all the thermodynamic observables follow. However, since an entropy measure undergoing the MaxEnt procedure cannot yield such a factorizable normalized structure if it has a non-additive slope, one cannot obtain the partition function and all related observables in a consistent manner.

We also numerically tested and verified our theoretical analysis for the Shannon, R\'enyi and Tsallis entropies. As can be seen in Figs. 1 and 2, the only factorizable normalized structure stems from the Shannon entropy, since its slope is additive. The  R\'enyi and Tsallis entropies yield this same result only when their deformation index $q$ becomes unity, thereby reducing to the Shannon measure.

There seems to be at least two main routes to follow if one wants to obtain factorized generalized equilibrium distributions: the first route is to modify the Lagrange multipliers which leads to contradiction as we have already shown. The second route is to consider different averaging schemes such as the escort averaging procedure for example. This second possibility requires a very detailed and non-trivial discussion by itself so that it will be presented elsewhere.

A word of caution is also in order: one might consider that the concomitant Legendre structure is preserved even for generalized entropies with non-additive slope such as Tsallis entropy, for example. Our work does not deny this fact, since we are here interested in whether the partition function used in the assessment of the Legendre structure is indeed a partition function factorized from the normalized optimal distribution obtained through the MaxEnt. Whether one has the factorized and normalized optimal distribution apparently precedes the issue of the preservation of the Legendre structure.

It is finally worth noting that for obtaining the canonical partition function, the current criterion sets more bounds on the respective entropy  measures compatible with the MaxEnt procedure than the Shore and Johnson axioms \cite{ShoreJ,Presse2}, since these axioms do not limit the use of the R\'enyi entropy in the context of MaxEnt \cite{Uffink} whereas the criterion of the additive slope does.

%============================


\begin{thebibliography}{0}
\bibitem{note}
Considering a two-state system ($n=2$), Eq. (\ref{eq:04}) is always solvable for any explicit expression of $I(p_1,p_2)$, since
%
\begin{eqnarray*}
p^\star_1&=&\frac{\varepsilon_2-U}{\varepsilon_2-\varepsilon_1}\,,\quad
p^\star_2=\frac{U-\varepsilon_1}{\varepsilon_2-\varepsilon_1}\,,\\
\alpha^\star&=&\frac{\varepsilon_2}{\varepsilon_2 - \varepsilon_1} f\rbr{\frac{\varepsilon_2-U}{\varepsilon_2-\varepsilon_1}} - 
\frac{\varepsilon_1}{\varepsilon_2 - \varepsilon_1} f\rbr{\frac{U-\varepsilon_1}{\varepsilon_2-\varepsilon_1}}\,,\\
\beta^\star&=&\frac{1}{\varepsilon_2 - \varepsilon_1} \sqbr{
f\rbr{\frac{U-\varepsilon_1}{\varepsilon_2-\varepsilon_1}} - f\rbr{\frac{\varepsilon_2-U}{\varepsilon_2-\varepsilon_1}}}.
\end{eqnarray*}
%
This is a very particular case, in which the MaxEnt distribution $p^\star_i$ is independent from the entropy measure.
\end{thebibliography}
\end{document}